\documentstyle[prb,twocolumn,aps,epsf,epsfig]{revtex}
\begin{document}
\draft
\twocolumn[\hsize\textwidth\columnwidth\hsize\csname@twocolumnfalse%
\endcsname

\title{
Manifestation of quantum chaos on scattering techniques:
application to low-energy and photo-electron diffraction
intensities.
}

\author{
P.L. de Andres and J.A. Verg\'es
}

\address{
Instituto de Ciencia de Materiales, Consejo Superior de 
Investigaciones Cient\'{\i}ficas,
Cantoblanco,
E-28049 Madrid, Spain} 

\date{\today}

\maketitle

\begin{abstract}
\baselineskip=2.5ex
Intensities of LEED and PED are analyzed
from a statistical point of view. The
probability distribution is compared with
a Porter-Thomas law, characteristic of 
a chaotic quantum system. The agreement obtained
is understood in terms of analogies between
simple models and Berry's conjecture for
a typical wavefunction of a chaotic system.
The consequences of this behaviour on
surface structural analysis are qualitatively 
discussed by looking at the behaviour of standard 
correlation factors.
\end{abstract}


\narrowtext

There is a continuous interest in understanding the
relationship between chaos and quantum mechanics.
Long time ago, Wigner
investigated 
the influence of chaos on quantum mechanical scattering 
experiments in nuclear systems.\cite{wigner}
Thereafter, much work has concentrated on
the analysis of energy levels of bound states
inside closed systems 
(like various types of billiard geometries).
While these studies offer obvious advantages,
a great deal of information is lost 
by neglecting the examination of
the wavefunctions.
In fact, a good understanding of
wavefunctions is crucial to explain
open systems, like
the standard probe-target-detector
setup used in most scattering experiments.
Therefore, it is quite perplexing to find so
few examples in the literature
related to quantum chaos manifested in experiments
where wavefunctions are analyzed,
which should be emphasized,
only correspond 
to closed geometries.\cite{alt,wilkinson}
In this work, we
show that
reflected intensities of surface
scattering experiments, which are 
directly related to the
modulus squared of the wavefunction,
are consistent with
quantum chaos.
Therefore, we are proposing 
a new class of 
simple experimental systems
where quantum chaos
is manifested in the properties of
wavefunctions.

One reason to expect quantum chaotic behaviour
in a scattering experiment
comes from the existence 
of classical chaos when three or more
scattering potentials 
are involved.\cite{gutzwiller}
Actually,
Mucciolo et al.\cite{mucciolo} have recently shown that 
the high energy region of 
the calculated
band structure of Si and ${\rm Al_{x}Ga_{1-x}As}$
is complex enough to obey the statistical
distribution of levels corresponding 
to random matrix theory (RMT).\cite{mehta}
Based on this statistical analysis,
the authors claim that these systems
exhibit quantum chaos.
Although just a theoretical prediction,
this is a remarkable result because no
disorder or incommensurate geometries
are involved, 
and 
the physical reason for chaos should be
found elsewhere (e.g., the intrinsic
multiple scattering (MS) originating Bloch
states and giving rise to the crystal band
structure).

In this letter, 
the manifestation of chaos on
standard surface scattering techniques 
like 
Low Energy Electron Diffraction (LEED)
or Photoelectron Diffraction (PED)
is investigated.
These techniques are dominated in most experimental
systems by MS,
yielding a clear similitude to the band
structure problem mentioned above.
In surface structural work, it is a common belief that MS 
introduces a richier but more difficult analysis.
This general statement is analyzed here 
from the point of view of the chaotic component 
of the LEED and PED experiments,
while we
draw new attention to a class of
quantum systems obeying a Porter-Thomas
probability distribution.\cite{porter}
Ultimately, our aim is the
understanding of the relationship 
between complex scattering phenomena
and the emergence of quantum chaos.

To characterize chaotic wavefunctions
Porter and Thomas\cite{porter} advanced the hypothesis
that wavefunctions of a
chaotic system should display a 
$\chi^2_{\nu}$
statistical probability
distribution.
Subsequently, this hypothesis 
has been rigorously justified 
using the supersymmetry formalism,\cite{efetov}
and has been used as a convenient definition of quantum chaos,
that at least can be thought as a
necessary condition.
Dyson\cite{dyson} demonstrated that within
the RMT only three universal classes
can exist depending on whether 
the hamiltonian is constructed with
real numbers, complex numbers or quaternions,
corresponding respectively to $\nu=1,2$, and $4$
degrees of freedom.
Since scattering wavefunctions are
complex numbers, the statistics corresponding to
the universality class
$\nu=2$ is expected.

An interesting theoretical result on the wavefunction
of a chaotic system
is due to Berry.\cite{berry}
Analysing the semiclassical mechanics of regular
and irregular motion, he realized that
a typical chaotic wavefunction
should be a linear combination
of plane waves with random $\vec k$-orientations,
(at a fixed constant energy), and
random complex coefficients:

\begin{equation}
\psi_{k}(\vec x) \propto 
\sum_{j=1,N} a_{j} e^{i \delta_{j}}
e^{i \vec k_{j} . \vec x}
\end{equation}

\noindent
Heller et al.\cite{berry} have further investigated the
properties of these chaotic wavefunctions,
finding that in 2D they present characteristic scars.
Berry's chaotic wavefunction can be interpreted
as the result of multiple random reflections
of a plane wave.
It can also be thought as
the superposition of plane waves 
originating at $N$ points 
propagating
with the same energy in random orientations,
and mixed with appropriated coefficients.
Guided by these images, we try to find 
a physical system where a similar wavefunction
can be realized. 

First of all, we consider a PED
experiment where an electron inside an atomic core 
is excited by an incident X-ray photon,
to be subsequently diffracted by a
cluster of $n$ atoms surrounding the
original source. 
This is a complicated scenario,
but making some approximations
that have been tested 
in the literature,
a simple expression
for the wavefield
at a distance $R$ (far-field) is:\cite{leependry}

\begin{equation}
\psi_{k}(\hat{k}) =
{ e^{i k R} \over i k R } \Big( 
1 + 
\sum_{\omega=1}^{n+{n \choose 2} + ...} 
c_{\vec r_{\omega}} 
e^{-i \vec k \vec r_{\omega}} \Big) ~~~~~~~~~,
\end{equation}

\noindent
where 
the complex coefficients, $c_{\vec r_{\omega}}$, include
appropriated scattering factors 
and the expansion can be extended to any desired
order of scattering.  
It is important to
realize the similarity of this MS 
series with Berry's one: 
taking away the prefactor and
the source wave, 
and given a fixed direction in
real space determined by $\hat{k}$,
it is written as a sum
of $N$ plane waves $e^{i \vec k . \vec r_{\omega}}$
with complex
coefficients, 
where the many $\vec r_{\omega}$ may result 
oriented in
uncorrelated directions if
enough scattering is allowed.
The following question arises in this context:
how many plane waves are necessary to allow 
Eq. (1) to follow a Porter-Thomas statistical
distribution? As an example, if
$k = 2 \pi$, we find that
just $N=10$ are enough to
find a reasonable agreement (e.g., see Fig. 1);
the result for as few as $N=2$ is
given in the same figure for comparison.

Secondly, we notice that 
an expression that is formally similar
to Eq. (2) 
can be written
for the Diffuse LEED (DLEED) wavefield:\cite{saldin}

\begin{equation}
\psi_{k} (\hat{k}) = 
F_{0}(\vec k) 
e^{i \vec k . \vec r} +
\sum_{\alpha} F_{\alpha}(\vec k) 
e^{i \vec k .( \vec r - \vec r_{\alpha})} ~~~~~~~~~~,
\end{equation} 

\noindent
where $F_{\alpha}$ represent generalized scattering
factors including MS contributions.
Standard LEED I(V) curves can be described in the same way,
for a given energy,
just keeping in mind that if the system
exhibits perfect periodicity in the parallel
direction, only a discrete set of points
given by Bragg conditions are available.

Before trying to analyze
real experiments, 
a set of
controlled theoretical simulations of
relevant systems is considered. 
We investigate the behaviour
of the single scattering term in Eq. (2) performing
the summation over a set of 500 atoms randomly distributed
around the origin between $r_{\alpha}=10$ a.u. and
$r_{\alpha}=150$ a.u.
(atomic units will be used throughout the paper,
expressing distances in Bohrs and energies
in Hartrees).
The central region (magnified ten times)
of a typical $\mid \psi_{k} \mid^{2}$ ($k=6$ a.u.)
measured on a sphere at an
asymptotic distance is shown in Fig. (2).
A typical {\it worm-like} image is
obtained when the pattern saturates at high
energies or high $r_{\alpha}$ distances.
The corresponding probability
distribution of intensities
(normalized to the average)
is seen to follow a $\chi^{2}_{2}$ distribution rather
well.
 
Using the same model,
we explore the probability distribution
of intensities produced by
a small quasi-regular cluster of
atoms. A cube
of $2 \times 2 \times 2$ Ni atoms centered
around the origin at a distance of $5$ a.u
is chosen.
Phase shifts up to $l_{max}=7$ are
used to compute the scattering factors.
These phase shifts have been obtained from a
muffin-tin model,
and can be used to represent
the scattering of electrons by atoms in
the real structure. 
In order to simulate small geometrical
irregularities caused by relaxations,
reconstructions or simply the effect
of temperature, the atoms
are randomly displaced from their perfect positions
in the cube
with values uniformly
distributed between $0$ and $\sqrt{3} d$. 
Fig. 1 shows the result of
such a simulation for $d=1$ a.u..
The agreement with the
Porter-Thomas law is really good. In the same
figure, the effect of
a smaller arbitrary displacement
($d=0.2$ a.u.) is also shown, together
with a similar calculation
for a double-scattering
term in a cubic array of
$3 \times 3 \times 3$ Ni atoms
with comparable results.
The analysis of Eq. (1) proves that fluctuations
are responsible for the appearance of
the ideal Porter-Thomas distribution.
Therefore, the deviations 
from the $\chi_{2}^{2}$ distribution
observed for the
scattering series should be
explained by studying their fluctuations, which 
is beyond the scope of this work.  
 
We compute
full dynamical diffuse Leed intensities\cite{cpcdleed}
for a realistic
adsorption geometry on the system ${\rm O/Ni(100)}$ (oxygen
is placed on the hollow site at $1.51$ a.u. from the
surface). All parameters are taken from 
a detailed structural analysis of the same system.\cite{uli}
It is worthwhile
to notice two points: 
i) All the Ni atoms
are placed at ideal bulk-like positions. 
Therefore, there
is not geometrical disorder in the problem,
the main source of complexity being MS
by the atoms in the ordered lattice;
and,
ii) All the calculations are performed at
$T=0$ K,
although attenuation effects are taken
into account in this formalism via an imaginary part
($V_{0i}= 0.15$ a.u.)
added to the energy.
Fig. 3 shows the probability distribution
for intensities
calculated theoretically at 
three different energies
($12$, $14$ and $16$ a.u.),
yielding a similar agreement to
the Porter-Thomas probability distribution
as the previous PED example.
The same results are expected from the analysis of
experimental intensities.
As an example, Fig. 3 includes a single
energy ($11.1$ a.u.) extracted from
the experimental database measured
by the Erlangen group.\cite{klausexp}

The same Porter-Thomas
probability distribution should also appear when
conventional LEED I(V) curves are analyzed, because
our arguments above   
are valid for 
{\it any} energy.
We have simulated theoretically the
LEED I(V) curves\cite{leed2} for three
materials with very different structures:
Cu(100), W(100), and Si(111).
An arbitrary non-normal incidence angle
($\theta=20^{\circ}$, $\phi=30^{\circ}$)
breaking the symmetry is chosen. This yields 
the maximum number of inequivalent beams
increasing the statistical confidence of the
analysis.
The first 9 emergent beams are used for Cu and W
while the first 13 were chosen for Si.
The energy ranged from 50 to 450 eV
for both metals and from 30 to 300 eV
for the semiconductor, yielding a
database of about $100$ a.u.,
which is enough for our purposes and
easily accessible to experiment.
The imaginary part of the energy
is fixed to a constant value of $0.15$ a.u.,
and 
$T=0$ K is used again.
Finally, we analyze the experimental 
database for $c(8 \times 2)$ ${\rm GaAs(100)}$,\cite{palomares}
formed by 13 different beams measured
at normal incidence.
Our results are shown in Fig. 4, displaying
an agreement with the Porter-Thomas
probability distribution similar to the 
other examples.

Guided by these results, we predict the existence
of a region (II) in parameter space, $P$, where small changes
in $\vec p$ (each component defining a relevant
parameter for the structure), result in rapid changes
of the wavefunction. On intuitive grounds, it can be
assumed that these changes must separate exponentially. 
This region is intermediate between
two different ones: a perturbative region (I), as
for sufficiently small changes in $\vec p$ we
expect perturbation theory to give a reasonable
answer,\cite{tleed} and a random region (III)
where wavefunctions for different structures are
uncorrelated. The existence of these
three regions is checked by analyzing 
two R-factors commonly used in
surface structure analysis: (i) A root mean square
displacement\cite{leed2} adapted for DLEED ($R_{2}$)
and, (ii) the Pendry R-factor\cite{rpendry} ($R_{P}$)
often used with standard LEED.

We apply $R_{2}$ to compare theoretically calculated
DLEED intensities for ${\rm O/Ni(100)}$ as a function of 
the adsorption height $h$. 
Arbitrarily, $h=1.51$ a.u. is chosen as the reference.
This is a clean and controlled theoretical experiment
where only the position of one atom is changed,
but to stay closer to reality, we also consider
the behaviour of the $R_{P}$ in a recent 
structural analysis of 
${\rm c}(2 \times 2) {\rm Si/Cu(110)}$ I(V) curves,
where the relaxation of the whole surface layer
is considered.\cite{celia}
Both cases show the existence of the three
regions mentioned above. Region I is obviously 
well characterized by the existence of a minimum
that imposes a quadratic dependence. Region III
is also easily identified by the saturation of 
the R-factor: for $R_{P}$ this happens 
by construction around $1$ (the maximum
value for $R_{P}$ can be twice this value,
but saturation starts at values
greater than 0.6). For
$R_{2}$ we observe that
saturation occurs around the value obtained
by comparing two sets of $N$ random intensities,
so the R-factor is normalized to this value.
Region II 
may be characterized
by plotting the $\ln (R)$ versus
a relevant component of $\vec p$, and identifying the
interval where it behaves like a 
straight line. 
When we use these ideas to analyze the data,
we find that the quadratic region (I) extends
approximately $0.04$ \AA\ for the DLEED case 
($R_{2} \le 0.2$) and 
$0.05$ \AA\ for the conventional LEED analysis
($R_{P} \le 0.25$).
The exponential region (II) extends also $0.04$ \AA\ for
DLEED ($R_{2} \le 0.5$), while it goes to $0.09$ \AA\ for the
LEED experiment ($R_{P} \le 0.6$).
Finally, an uncorrelated region
extends beyond these intervals, provided we
do not approach a multiple coincidence minima.
We notice that a perturbative technique where
the perturbation in the potential is proportional to the
atomic displacements (the so-called
Tensor LEED first approximation\cite{tleed})
is known to break down beyond
$\approx 0.1$ \AA.
This is close to regions I and II,
considered respectively a truly perturbative
region (quadratic)
and the onset of the breakdown for the perturbative
approach (exponential).
These findings should bring more
rigour to the standard R-factor analysis because
our analysis allow to identify regions II and III,
where correlations must be taken as spurious,
and offer a new theoretical explanation for the definitive
failure of simple scattering methods in LEED.

We have analyzed typical wavefunctions for LEED and
PED experiments in the light of Berry's proposal for
a generic chaotic wavefunction. 
Our statistical analysis shows that 
scattering wavefunctions computed from several models
(including perfectly ordered structures)
follow the Porter-Thomas $\chi^{2}_{2}$
distribution. This property is
also obtained analyzing 
experimental data for LEED and DLEED.
The physical origin of this behaviour
is the complexity of the scattering.
Attenuation effects 
taken into account via a complex optical
potential fitted to experiments,
and defects (relaxations or reconstructions)
do not change this conclusion.
Finally, analysing the behaviour of two
different R-factors, we have
argued the existence of three
distinct regions, showing the rationale behind
widely used rules about
which R-factors are acceptable 
in standard structural work and which 
are not.

We are grateful to Prof. K. Heinz
for making available to us his
experimental DLEED data on ${\rm O/Ni(100)}$.
This work has been supported by the CICYT under
contracts num. PB94-053 and MAT94-0058.

\begin{figure}
\epsfig{figure=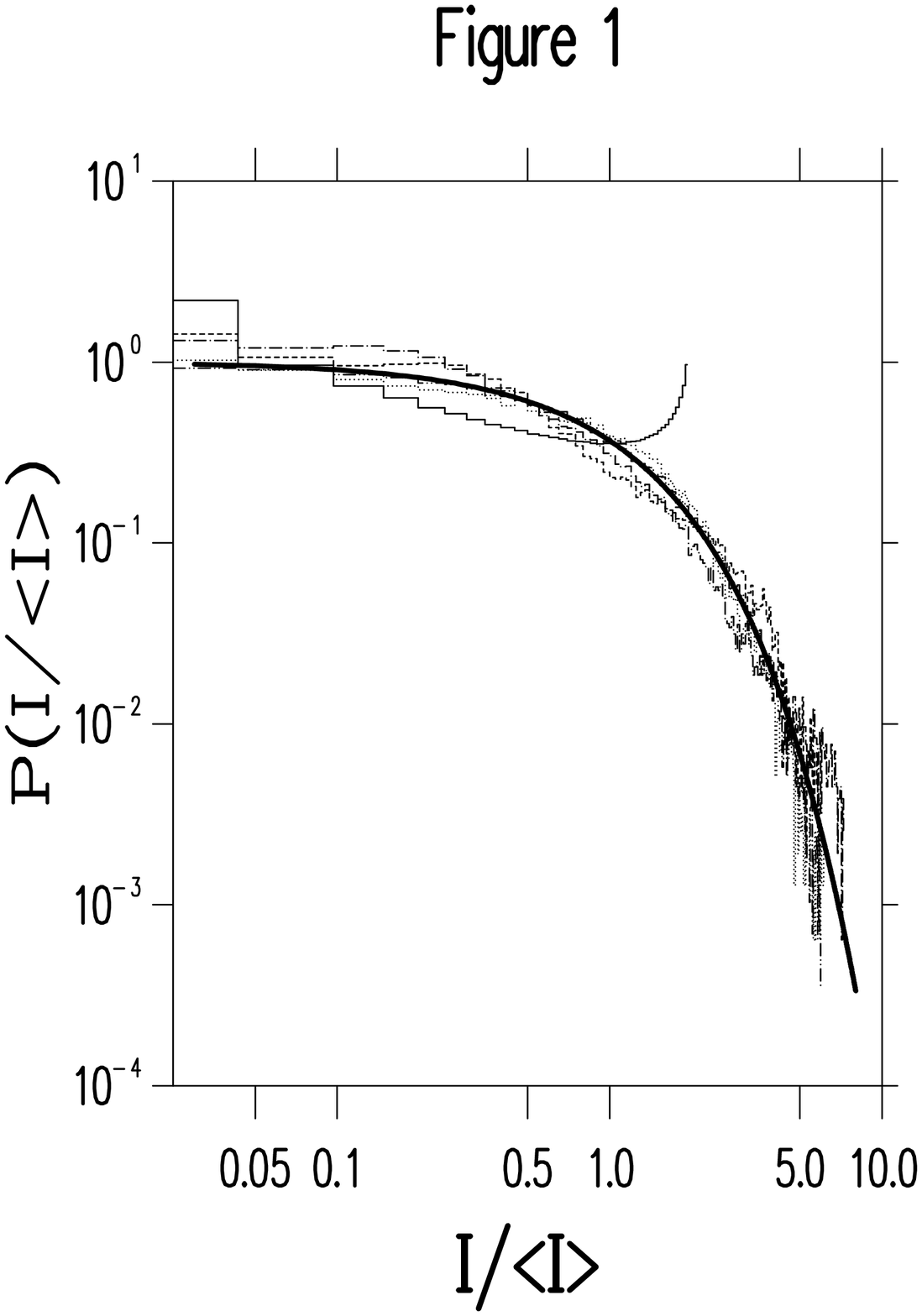,height=10cm,width=8.5cm}
\caption{
Probability distribution for simplified
models:
(a) Thick solid line: $\chi^{2}_{2}$ law;
(b) Thin solid line: Berry's wavefunction with $N=2$;
(c) dashed-dotted line: Berry's wavefunction with $N=10$;
(d) dotted line: simple PED model for eight Ni atoms
on a cube, simple
scattering ($E=1.8$ a.u., $d=1.0$ a.u.);
(e) dashed line: same as (d) with $d=0.2$ a.u.;
(f) dashed two-dotted line: simple PED model for 27 Ni
atoms, double scattering ($E=1.8$ a.u., $d=1.0$ a.u.).
}
\end{figure}

\begin{figure}
\epsfig{figure=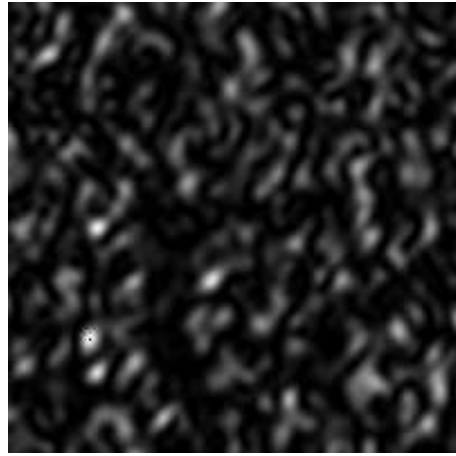,height=6.0cm,width=6.0cm}
\caption{
Small portion of a wavefunction (modulus squared)
obtained using 
Eq. (2) for a set
of 500 atoms at random positions.
}
\end{figure}

\begin{figure}
\epsfig{figure=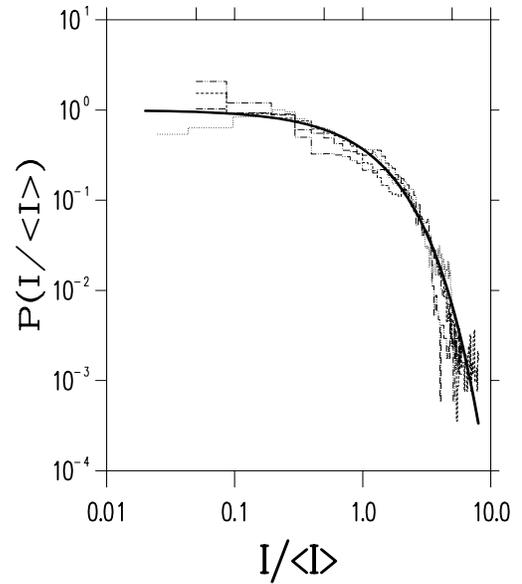,height=10cm,width=8.5cm}
\caption{
Probability distribution for MS
intensities in a DLEED
model.
Different energies are shown.
(a) Thick solid line: $\chi^{2}_{2}$ law;
(b) dotted: $E=12$ a.u.;
(c) dashed: $E=14$ a.u.;
(d) dashed dotted: $E=16$ a.u.;
(e) dashed two-dotted: $E=11.1$ a.u. (experimental).
}
\end{figure}

\begin{figure}
\epsfig{figure=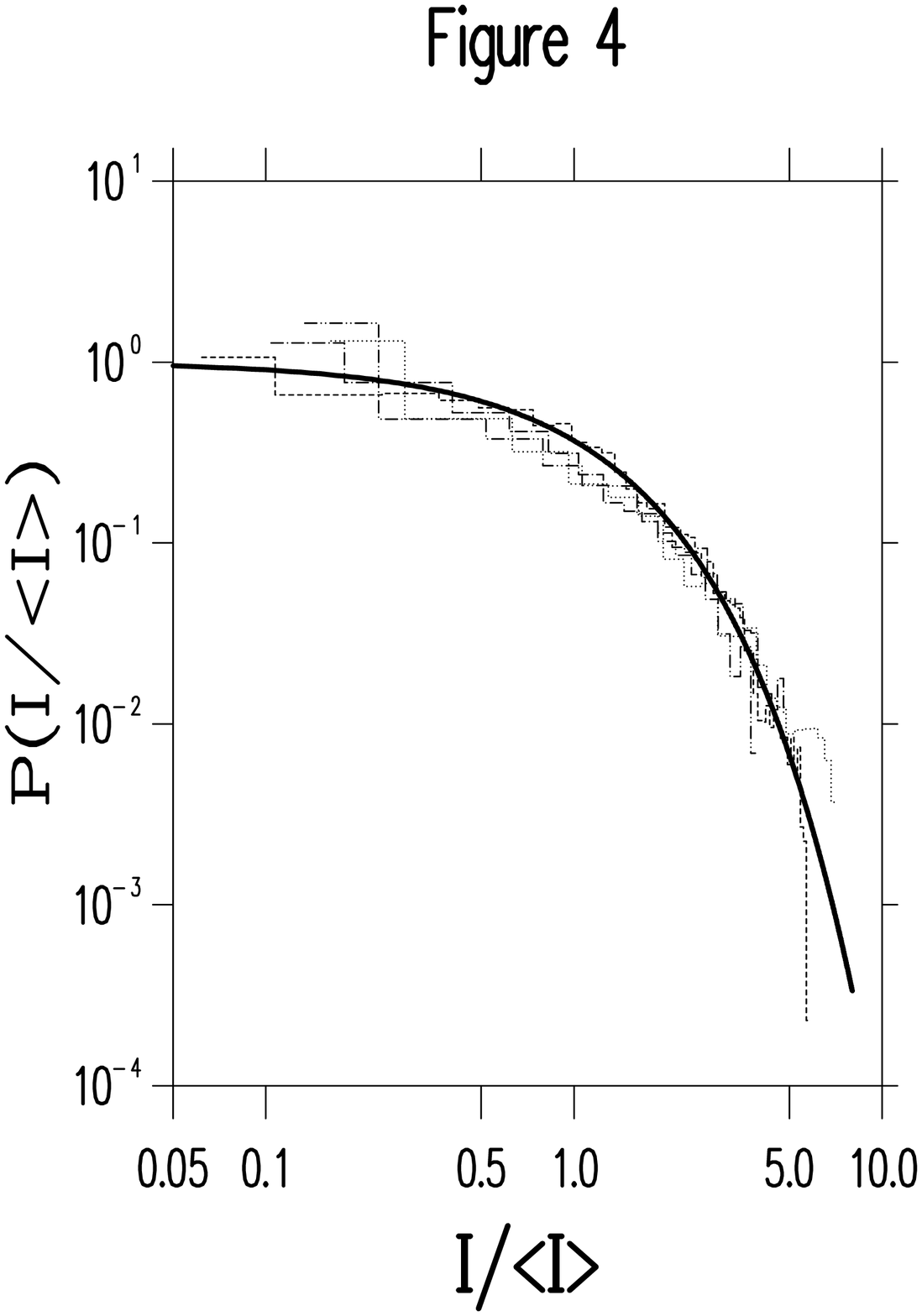,height=10cm,width=8.5cm}
\caption{
Probability distribution for MS
intensities in a LEED
model. Different materials are shown.
(a) Thick solid line: Porter-Thomas law;
(b) dotted: Cu(100);
(c) dashed: W(100);
(d) dashed dotted: Si(111);
(e) dashed two-dotted: 
${\rm c}(8 \times 2) {\rm GaAs(100)}$ (experimental).
}
\end{figure}


\begin{thebibliography}{10}

\bibitem[1]{wigner} E.P. Wigner, e.g. see
{\it Statistical Theories of Spectra:
Fluctuations}, Ed. by C.E. Porter,
Academic Press (New York, 1965).

\bibitem[2]{alt} H. Alt, H.-D. Gr\"af, H.L. Harney,
R. Hofferbert, H. Lengeler, A. Ritchter,
P. Schardt, V.A. Weidenm\"uller,
Phys. Rev. Lett. {\bf 74}, 62 (1995). 

\bibitem[3]{wilkinson} P.B. Wilkinson et al.,
Nature {\bf 380}, 608 (1996).

\bibitem[4]{gutzwiller} M.C. Gutzwiller,
{\it Chaos in Classical and Quantum Mechanics},
Springer Verlag (New York, 1990).

\bibitem[5]{mucciolo} E.R. Mucciolo, R.B. Capaz, 
B.L. Altshuler, J.D. Joannopoulos,
Phys. Rev. B. {\bf 50}, 8245 (1994). 

\bibitem[6]{mehta} M.L. Mehta, {\it Random Matrices},
2nd ed. (Academic Press, San Diego, CA, 1991).

\bibitem[7]{porter} C.E. Porter, R.G. Thomas, Phys. Rev.
{\bf 104}, 483 (1956).

\bibitem[8]{efetov} K.B. Efetov, V.N. Prigodin,
Phys. Rev. Lett. {\bf 70}, 1315 (1993).

\bibitem[9]{dyson} F.J. Dyson, J. Math. Phys.
{\bf 3}, 140 (1962).

\bibitem[10]{berry} M.V. Berry, in {\it Chaotic
Behaviour of Deterministic Systems}, Ed. by
G. Looss, R. Helleman and R. Stora (North Holland,
NY 1983), p. 171;
P.O'Connor, J. Gehlen, E.J. Heller,
Phys. Rev. Lett. {\bf 58}, 1296 (1987). 

\bibitem[11]{leependry} P.A. Lee and J.B. Pendry, Phys. Rev. B
{\bf 11}, 2795 (1975);
J.J. Barton and D.A. Shirley,
Phys. Rev. B {\bf 32}, 1906 (1985)

\bibitem[12]{saldin} D.K. Saldin and P.L. de Andres,
Phys. Rev. Lett. {\bf 64}, 1270 (1990).

\bibitem[13]{cpcdleed} J.B. Pendry and D.K.
Saldin, Surf. Sci. {\bf 145}, 33 (1984);
D.K. Saldin and J.B. Pendry, Comput. Phys.
Commun. {\bf 42}, 399 (1986).

\bibitem[14]{uli} U. Starke, P.L. de Andres, D.K. Saldin,
K. Heinz and J.B. Pendry, Phys. Rev. B, {\bf 38},
12277 (1988).

\bibitem[15]{klausexp} K. Heinz, private communication. 

\bibitem[16]{palomares} F.J. Palomares, Ph.D. thesis,
Chap. 3 (pg. 66),
Universidad Autonoma de Madrid, Madrid (1993).

\bibitem[17]{tleed} P.J. Rous, J.B. Pendry, D.K. Saldin,
K. Heinz, K. M\"uller, N. Bickel, Phys. Rev. Lett.
{\bf 57} 2951 (1986);
P.J. Rous, Ph.D. thesis,
Chap. 6 (pg. 167),
Imperial College of Science, Technology and
Medicine, London (1986).

\bibitem[18]{leed2} M.A. Van Hove and S.Y. Tong,
{\it Surface Crystallography by LEED}, (Springer-Verlag, Berlin,
1979);
M.A. Van Hove, W.H. Weinberg and C.M. Chan,
{\it Low-Energy Electron Diffraction},
(Springer-Verlag, Berlin, 1986).

\bibitem[19]{rpendry} J.B. Pendry, J. Phys. C: Solid State Phys.
{\bf 13} (1980) 937.

\bibitem[20]{celia} C. Polop, et al.,
unpublished.

\end{thebibliography}
\end{document}